\documentclass[11pt,a4paper]{article}
\usepackage{amsmath,amssymb,bm,epsfig,color,graphicx,cite,braket,mathrsfs,slashed,multirow}

\renewcommand{\thefootnote}{\fnsymbol{footnote}}

\usepackage{hyperref,xcolor}
\hypersetup{
    colorlinks=true,
    linktocpage=true,
    linkcolor=blue,
    filecolor=blue,      
    urlcolor=blue,
    citecolor=blue,
    }
\usepackage{siunitx}
\begin{document}

\title{
\begin{flushright}
\begin{minipage}{0.2\linewidth}
\normalsize
CTPU-PTC-22-14 \\*[50pt]
\end{minipage}
\end{flushright}
{\Large \bf 
Neutrino lines from MeV dark matter annihilation and decay 
\\ in JUNO
\\*[20pt]}}

\author{
Kensuke Akita$^{1}$\footnote{
E-mail address: \href{mailto:kensuke8a1@ibs.re.kr}{kensuke8a1@ibs.re.kr}},~
Gaetano Lambiase$^{2,3}$\footnote{
E-mail address: \href{mailto:lambiase@sa.infn.it}{lambiase@sa.infn.it}},~
Michiru Niibo$^{4,5}$\footnote{
E-mail address: \href{mailto:g2140626@edu.cc.ocha.ac.jp}{g2140626@edu.cc.ocha.ac.jp}},~
  and\
Masahide Yamaguchi$^{5}$\footnote{
E-mail address: \href{mailto:gucci@phys.titech.ac.jp}{gucci@phys.titech.ac.jp}}\\*[20pt]
$^1${\it \small
Center for Theoretical Physics of the Universe, Institute for Basic Science,
Daejeon 34126, Korea} \\
$^2${\it \small
INFN Sezione di Napoli, Gruppo collegato di Salerno, I-84084 Fisciano (SA), Italy} \\
$^3${\it \small
Dipartimento di Fisica "E.R. Caianiello", Universit\`a di Salerno, I-84084 Fisciano (SA), Italy} \\
$^4${\it \small
Department of Physics, Graduate School of Humanities and Sciences,}\\{\it\small  Ochanomizu University, Tokyo 112-8610, Japan} \\
$^5${\it \small
Department of Physics, Tokyo Institute of Technology,
Tokyo 152-8551, Japan} \\*[50pt]
}

\date{
\centerline{\small \bf Abstract}
\begin{minipage}{0.9\linewidth}
\medskip \medskip \small 
We discuss the discovery potential of JUNO experiment for neutrino lines from MeV dark matter (DM) annihilation and decay in a model-independent way. We find that JUNO will be able to give severe constraints on the cross section of DM annihilating into neutrinos and on the lifetime of DM decaying into neutrinos. More concretely, with $20$ years of data-taking in the fiducial volume $17$ kton, the cross section will be constrained smaller than $4\times 10^{-26}\,{\rm cm^{3}\,sec^{-1}}$ for the mass of a DM particle $15\,{\rm MeV} \lesssim m_{\chi} \lesssim 50\,{\rm MeV}$ at $90\,\%$ C.L., which might be strong enough to test thermal production mechanism of DM particles for such range of DM mass. The lifetime will be constrained as strong as $1\times 10^{24}\,{\rm sec}$ for the mass of a DM particle $m_{\chi} \simeq 100\,{\rm MeV}$ at $90\,\%$ C.L..
\end{minipage}
}

\maketitle{}
\thispagestyle{empty}
\clearpage
\tableofcontents
\clearpage

\renewcommand{\thefootnote}{\arabic{footnote}}
\setcounter{footnote}{0}

\section{Introduction}
\label{sec1}

The existence of dark matter (DM) is established by cosmological and astrophysical observations while the nature of dark matter particles such as their masses and their interactions with ordinary matter remains one of the most outstanding mysteries in particle physics and cosmology.

If DM is thermally produced through a weak interaction with the Standard Model (SM) particles, then DM has to annihilate in the early universe to account for the observed relic abundance. This thermal production mechanism relies on the assumption that DM can annihilate into the SM particles today. The observed relic abundance of DM requires the annihilation cross section of $\langle \sigma v \rangle \simeq (3$-$4) \times 10^{-26}\ {\rm cm^3\  sec^{-1}}$. A recent analysis constrains DM mass as $m_{\chi}\gtrsim 20\ {\rm GeV}$ for thermal relic DMs with a $2\rightarrow 2\ s$-wave annihilation process into the SM particle except for neutrinos \cite{Leane:2018kjk}. The model-independent lower bound on the mass of thermal DMs annihilating into neutrinos is $m_\chi \gtrsim 3.7\ {\rm MeV}$, obtained by measurements of the effective neutrino number $N_{\rm eff}$ from the Cosmic Microwave Background and the Big Bang Nucleosynthesis \cite{Boehm:2013jpa,Nollett:2013pwa,Nollett:2014lwa,Escudero:2018mvt,Sabti:2019mhn}.

An alternative but interesting possibility for DM to produce visible signals is DM decay. Since DM needs to survive until the current epoch of the universe, it has a lifetime at least longer than the age of the universe, $t_0\simeq 4.35\times 10^{17}\ {\rm sec}$. However, in principle, it can decay into lighter particles finally. The current model-independent and the severest constraints on DM lifetime give its lower bound around 250 Gyr \cite{Audren:2014bca,Enqvist:2019tsa,Nygaard:2020sow,Alvi:2022aam,Simon:2022ftd}. If DM would decay into the SM particles except for neutrinos, the observations of isotropic $\gamma$-ray background impose a lower limit of $\tau_\chi\sim (1\textrm{-}5)\times 10^{28}\ {\rm sec}$ for DM masses of $10\ {\rm GeV}$ to $1\ {\rm ZeV}$ \cite{Blanco:2018esa}.

Among the final states of the SM particles from DM, neutrino signals are the most difficult to capture.
However, in principle, we can observe such neutrino signals from DM through the charged leptons via the charged-current weak interactions. In addition, several models of DM producing a neutrino signal have been already proposed.
For example, the model of DM decay into neutrinos is proposed as the Majoron dark matter, e.g.\ in Refs.~\cite{Rothstein:1992rh,Berezinsky:1993fm}. In this model, spontaneous breaking of global $U(1)_{\rm B-L}$ symmetry in the SM, which realizes the seesaw mechanism \cite{Minkowski:1977sc,Yanagida:1979as,Gell-Mann:1979vob} to account for the smallness of neutrino masses, generates a pseudo-Nambu-Goldstone boson called Majoron \cite{Chikashige:1980ui,Schechter:1981cv}. Since the couplings of the Majoron are suppressed by the $U(1)_{\rm B-L}$ breaking scale, i.e., the seesaw scale, the Majoron can be dark matter with the lifetime much longer than the age of the universe, which is consistent with the current constraints on the lifetime of Majoron DM \cite{Garcia-Cely:2017oco}.

Albeit the difficulties in detecting the neutrino signals, the neutrino detectors with very large volumes, such as Super-Kamiokande \cite{Palomares-Ruiz:2007trf}, IceCube, and ANTARES, have already set stringent limits on DM annihilation cross section and DM lifetime over certain mass scales of DM. However, current constraints on DM annihilation cross section into neutrinos have not yet reached $\langle \sigma v \rangle \simeq (3$-$4) \times 10^{-26}\ {\rm cm^3\  sec^{-1}}$ over almost all the mass scales of DM (see e.g. Fig.~2 of Ref.~\cite{Arguelles:2019ouk}). It is an interesting question whether future experiments will be able to prove or disprove the thermal production mechanism.
On the other hand, DM lifetime decaying into neutrinos has already been constrained to be much longer than the age of the universe through neutrino detection experiments like Super-Kamiokande \cite{Palomares-Ruiz:2007egs}. Such constraints are more severe than those coming from the Cosmic Microwave Background(CMB) for some region of mass scales. 

In the near future, the next generation of large neutrino detectors, such as Hyper-Kamiokande (HK) \cite{Hyper-Kamiokande:2018ofw}, JUNO \cite{JUNO:2015zny}, DUNE \cite{DUNE:2020ypp}, IceCube-Gen2 \cite{IceCube-Gen2:2020qha} and KM3NET \cite{KM3Net:2016zxf}, are expected to improve their sensitivities, which will place more stringent constraints on the neutrino signal from DM. 
In fact, HK is expected to probe the annihilation cross section into neutrinos of $4\times 10^{-26}\, {\rm cm^3}\, {\rm sec^{-1}}$ for DM mass around $20\, {\rm MeV}$, depending on the DM halo profile \cite{Olivares-DelCampo:2018pdl,Bell:2020rkw}. The upcoming CMB-S4 observation is also expected to probe this annihilation cross section for DM mass up to $10$-$15\, {\rm MeV}$ \cite{Escudero:2018mvt,Sabti:2019mhn}.

The Jiangmen Underground Neutrino Observatory (JUNO) \cite{JUNO:2015zny} is a 20 kton neutrino detector made of linear alkylbenzene liquid scintillator ($\mathrm{C_6H_5C_{12}H_{25}}$) to be built in Jiangmen, China. Though the detector volume is 20 kton comparable to Super-Kamiokande, JUNO has very good energy resolution because of the liquid scintillator instead of water, which makes JUNO very sensitive to monochromatic signals. DM annihilation and decay inside the Milky Way can produce monochromatic neutrino lines while DM outside the Milky Way can produce diffuse signals owing to cosmic expansion. Thanks to the high DM density in the Milky Way, the neutrino lines are more prominent than the diffuse neutrino signals. In particular, JUNO will detect neutrinos with MeV energy range by the inverse beta decay (IBD) process, $\bar{\nu}_e+p\rightarrow e^++n$, with $E_{e^+} \simeq E_{\nu}-1.3\ {\rm MeV}$\footnote{To be more precise, at higher energy of neutrinos, the correction to the positron energy is non-negligible, $\Delta E_{e^+}\simeq 2E_\nu^2/m_p$ where $\Delta E_{e^+}$ is the difference between the maximum and minimum positron energy and $m_p$ is the proton mass \cite{Palomares-Ruiz:2007trf}. We take into account this correction in the following estimations.},
where $E_{e^+}$ and $E_{\nu}$ are the positron and neutrino energies, respectively. Since the expected MeV-scale positron signals from DM inside the Milky Way in JUNO will be approximately monochromatic, JUNO will have very good potential for discovering neutrino lines from MeV DM annihilation and decay.

In this paper, we discuss the discovery potential of JUNO for neutrino lines from MeV DM decay as well as its annihilation in a model-independent way. Though a similar analysis has already been done in Refs.~\cite{Klop:2018ltd, Arguelles:2019ouk} for annihilation, we extend the analysis by taking into account longer data-taking, wider mass range, and different DM profiles. We find that JUNO will be able to place more severe constraints on both DM annihilation cross section and decay rate into neutrino lines than the current constraints at the MeV region of DM mass due to its excellent energy resolution.

This paper is organized as follows. In the next section, we briefly review basic formulae to estimate the neutrino fluxes coming from DM annihilation and DM decay. In Sec.~\ref{sec3}, we estimate event rates and backgrounds of anti-neutrino captures via the IBD process in JUNO experiment, and then discuss the discovery potential for such neutrino lines through a $\chi^2$ analysis. The final section is devoted to conclusions. In Appendix~\ref{appa}, we estimate event rates and backgrounds of anti-neutrino captures via the charged-current neutrino interaction with $^{12}\mathrm{C}$ in addition to the IBD process and discuss the discovery potential of JUNO experiment including such interaction.

\section{Neutrino flux from dark matter in the Milky Way }
\label{sec2}

In this section, we briefly review neutrino flux from dark matter annihilation and decay in the Milky Way. We do not consider the diffuse neutrino flux from DM annihilation and decay outside the Milky Way since the peak of this diffuse flux smeared by the cosmic expansion is much lower than that of the flux produced from the high DM density in the Milky Way.
On the other hand, as will be discussed in the next section, we will take the events of the diffuse supernova neutrino background (DSNB) into account as a background. Thus, our analysis would give conservative constraints.

\subsection{Dark matter annihilation}
\label{sec2.1}
In the following, we consider DM annihilation into one neutrino-antineutrino pair, $\chi\chi\rightarrow \nu \bar{\nu}$ for simplicity.
The differential neutrino flux per energy and flavor from the DM annihilation is given by
\begin{align}
    \frac{d\Phi_{\nu}}{dE_\nu}=
    \frac{\langle\sigma v\rangle}{2}\frac{\mathcal{J}}{4\pi m_\chi^{2}}\frac{1}{3}\frac{dN_\nu}{dE_\nu},\label{eq:DM_anni_flux}
\end{align}
where $m_{\chi}$ and $\langle \sigma v \rangle$ are the DM mass and the averaged annihilation cross section, respectively. The same formula of the differential flux holds for anti-neutrinos. Here we assume that DM is a Majorana particle. 
The factor $1/2$ accounts for the initial states of identical two Majorana DM particles.
For a Dirac DM particle, we need to replace the factor $1/2$ to $1/4$ by taking into account the difference between DM particles and anti-DM particles instead of 1/2 for the initial two identical Majorana DM. The factor of $1/3$ comes from our assumption that all flavor neutrinos are populated due to the same branching ratio of the DM annihilation for each flavor or due to neutrino mixing. $dN_\nu/dE_\nu$ is the neutrino energy spectrum produced by one DM annihilation.

Since the emission of scintillation light is isotropic in JUNO, 
we use the J-factor $\mathcal{J}$, defined by integrating the squared DM density over the line of sight $s$ and the target solid angle as follows, 
\begin{align}
    \mathcal{J}=\int_{0}^{2\pi}dl\int^{\pi/2}_{-\pi/2}db \cos{b} \int_0^{s_{\rm max}} ds\ \rho\left(\sqrt{R_{\rm sc}^2-2sR_{\rm sc}\cos\psi+s^2} \right)^{2},    \label{Jfactor}
\end{align}
where $(b,l)$ are galactic coordinate, $\cos{\psi} = \cos{b}\cos{l}$, $\rho(r)$ is the DM density in the Milky Way, $R_{\rm sc}\simeq8\ {\rm kpc}$ is the distance from the Solar System to the Galactic Center, 
and the upper limit of integration is $s_{\rm max} = \sqrt{R_{\rm MW}^2 - \sin^{2}{\psi}R_{\rm sc}^{2}} + R_{\rm sc}\cos{\psi}$, where $R_{\rm MW}=40\ {\rm kpc}$ is the size of the Milky Way.

The DM density $\rho(r)$ includes astrophysical uncertainties.
N-body simulations for the structure formation in the universe, which successfully reproduce the structure on large scales, can also predict the DM density precisely on such scales. However, the DM density in the innermost part of the galaxy is still under dispute. Integrating the DM density over the whole Milky Way, the uncertainties of the DM profiles included in the J-factor would affect less the neutrino flux from the DM annihilation \cite{Yuksel:2007ac}.
For reference, we adopt the following four profiles,
the generalized Navarro-Frenk-White (NFW) \cite{Benito:2019ngh}, NFW\cite{Navarro:1995iw}, Moore\cite{Moore:1999gc} and Isothermal\cite{1980ApJS...44...73B} profiles, which are parametrized as
\begin{align}
    \rho(r)=\frac{\rho_{0}}{\left(\frac{r}{r_s}\right)^{\gamma}\left[1+\left(\frac{r}{r_s}\right)^\alpha \right]^{(\beta-\gamma)/\alpha}},
    \label{DMprofile}
\end{align}
where $\alpha,\beta$ and $\gamma$ are slope parameters and $r_s$ is the scale radius. We assume the local density at the Solar System of $\rho(R_{\rm sc})\sim 0.3$-$0.4\ {\rm GeV\, cm^{-3}}$ for the four profiles. 
The parameters are shown in Table~\ref{tb:ParameterDMH}.

For the DM annihilation, $\chi\chi\rightarrow \nu\bar{\nu}$, the neutrino energy spectrum produced by one DM annihilation is given by
\begin{align}
    \frac{dN_\nu}{dE_\nu}=\delta \left(E_\nu-m_{\chi}\right).
\end{align}
We also have the same expression for anti-neutrinos.

\begin{table}[h]
\begin{center}
	\begin{tabular}{ccccccccc}
		\hline \hline
	     Profile & $\alpha$ & $\beta$ & $\gamma$ & $r_s$ & $R_{sc}$ & $\rho(R_{\rm sc})$ & $\mathcal{J}$ & $\mathcal{D}$ \\
		\hline 
		Generalized NFW & 1 & 3 & 1.2  & 20 & 8.127 & 0.4 & 2.3 & 2.5\\
		NFW & 1 & 3 & 1 & 20 & 8.5 & 0.3 & 0.88 & 1.9\\
		Moore & 1.5 & 3 & 1.5 & 28 & 8.5 & 0.27 & 2.7 & 2.5\\
		Isothermal & 2 & 2 & 0 & 5 & 8.5 & 0.3 & 0.53 & 1.9\\
		\hline \hline
	\end{tabular}
	\caption{The parameters of  Eq.~(\ref{DMprofile}) for the generalized NFW\cite{Benito:2019ngh}, NFW \cite{Navarro:1995iw}, Moore \cite{Moore:1999gc} and Isothermal \cite{1980ApJS...44...73B} halo profiles. The J-factors (\ref{Jfactor}) and the D-factors (\ref{Dfactor}) for the four profiles are also shown. The units for $r_s$, $R_{sc}$, $\rho(R_{\rm sc})$, $\mathcal{J}$ and $\mathcal{D}$ are $[\rm{kpc}]$, $[\rm{kpc}]$, $[{\rm GeV\, cm^{-3}}]$, $[10^{23}\, {\rm GeV^2\, cm^{-5}}]$ and $[10^{23}\, {\rm GeV\, cm^{-3}}]$, respectively.}
  \label{tb:ParameterDMH}
\end{center}
\end{table}
\subsection{Dark matter decay}
\label{sec2.2}

In the following, we consider DM decay into one neutrino-antineutrino pair, $\chi\rightarrow \nu \bar{\nu}$, for simplicity. If DM lifetime is longer than the age of the universe, i.e., the density of DM in the Milky Way does not change significantly due to its decay, the differential neutrino flux per energy and flavor is given by
\begin{align}
    \frac{d\Phi_{\nu}}{dE_\nu}=
    \frac{1}{\tau_\chi}\frac{\mathcal{D}}{4\pi m_\chi}\frac{1}{3}\frac{dN_\nu}{dE_\nu},\label{eq:DM_decay_flux}
\end{align}
where $m_{\chi}$ and $\tau_\chi$ are the DM mass and lifetime, respectively. We obtain the same differential flux for anti-neutrinos.
The factor of $1/3$ comes from our assumption that all flavor neutrinos are populated due to the same branching ratio of the DM decay for each flavor or due to neutrino mixing. $dN_\nu/dE_\nu$ is the neutrino energy spectrum produced by a DM decay.

We also use the D-factor integrated over angles of incoming anti-neutrinos for the case of JUNO. Such D-factor, $\mathcal{D}$, for the DM decay is obtained by integrating the DM density itself, rather than the square of it as in the case of the DM annihilation, over the line of sight $s$ and the target solid angle as
\begin{align}
    \mathcal{D}=\int_{0}^{2\pi}dl\int^{\pi/2}_{-\pi/2}db \cos{b} \int_0^{s_{\rm max}} ds\ \rho\left(\sqrt{R_{\rm sc}^2-2sR_{\rm sc}\cos\psi+s^2} \right),
    \label{Dfactor}
\end{align}
where $\cos{\psi} = \cos{b}\cos{l}$, $R_{\rm sc}\simeq 8\ {\rm kpc}$, and $s_{\rm max} = \sqrt{(R_{\rm MW}^2 - \sin^{2}{\psi}R_{\rm sc}^{2})} + R_{\rm sc}\cos{\psi}$ with $R_{\rm MW}=40\ {\rm kpc}$.

The neutrino flux from the DM decay depends on the line of sight integral of the DM density itself, rather than the square of it as in the case of the DM annihilation, so that the D-factor $\mathcal{D}$ is less affected by the profile uncertainties compared to the J-factor $\mathcal{J}$\cite{Palomares-Ruiz:2007egs}.
In addition, integrating the DM density over the whole Milky Way, the uncertainties of the DM profiles would not affect the neutrino flux from the DM decay significantly \cite{Yuksel:2007ac}.
For reference, we adopt the generalized NFW, NFW, Moore, and Isothermal profiles and show the D-factors for their profiles in Table~\ref{tb:ParameterDMH}.

For the DM decay, $\chi\rightarrow \nu\bar{\nu}$, the neutrino energy spectrum produced by one DM decay is given by
\begin{align}
    \frac{dN_\nu}{dE_\nu}=\delta \left(E_\nu-\frac{m_\chi}{2}\right).
\end{align}
We also have the same expression for anti-neutrinos.

\section{Signatures of MeV-scale neutrino lines in JUNO}
\label{sec3}

In this section, we estimate the discovery potential of JUNO for neutrino lines from the DM decay, $\chi\rightarrow \nu_e\bar{\nu}_e$, as well as from the DM annihilation, $\chi\chi\rightarrow \nu_e\bar{\nu}_e$.
We introduce relevant backgrounds in JUNO in Section~\ref{sec3.1}, and discuss the discovery potential of JUNO in Section~\ref{sec3.2}.

\subsection{Event rates and backgrounds}
\label{sec3.1}

The Jiangmen Underground Neutrino Observatory (JUNO) \cite{JUNO:2015zny} is a 20 kton neutrino detector made of linear alkylbenzene liquid scintillator ($\mathrm{C_6H_5C_{12}H_{25}}$) to be built in Jiangmen, China. An excellent energy resolution of JUNO will be achievable because of the liquid scintillator detector. 

The main detection channel in JUNO is the inverse beta decay (IBD) process, $\bar{\nu}_e + p \rightarrow e^+ + n$. In liquid scintillator detector, the neutron is captured on a free proton with a capture time of $\sim 200\, \mu s$, emitting a 2.2 MeV photon \cite{JUNO:2015zny},
Thus, the IBD process is a double coincidence signal event with a characteric time of $\sim 200\,  \mu s$ and 2.2 MeV photon, which may be useful for reducing accidental background\cite{JUNO:2015zny}.
The subdominant detection channel is the charged-current (CC) neutrino interactions with $\mathrm{^{12}C}$, $\bar{\nu}_e + ^{12}\mathrm{C}\rightarrow ^{12}\mathrm{B} + e^+$ and $\nu_e + ^{12}\mathrm{C} \rightarrow ^{12}\mathrm{N} + e^-$.
These processes are also double coincidence signal events due to the beta decays of $^{12}\mathrm{B}$ and $^{12}\mathrm{N}$ with a $20.2\, {\rm ms}$ and $11\, {\rm ms}$ half lifetime, respectively \cite{JUNO:2015zny}.
These different time delayed signals can help us to distinguish between the IBD process and the CC interactions with $\mathrm{^{12}C}$.
In the following, we neglect the subdominant processes of the CC neutrino interactions with $\mathrm{^{12}C}$. We will discuss the discovery potential of JUNO including the CC interactions with $\mathrm{^{12}C}$ in Appendix~\ref{appa}.

The relevant backgrounds at $E_{e^+}\lesssim 10\ {\rm MeV}$ are reactor neutrinos near JUNO. At $10\ {\rm MeV} \lesssim E_{e^+}\lesssim 30\ {\rm MeV}$, the main backgrounds are the events of atmospheric neutrinos through neutral current (NC) interactions and the events of the diffuse supernova neutrino background (DSNB), which is diffuse neutrinos produced from all past supernovae in the universe. The event of fast neutrons produced by decays of cosmic muon outside the detector also contributes to the background. When we consider a smaller fiducial volume of 17 kton, this background can be reduced since most of the fast neutron events happen near the edge of the detector. In addition, the pulse-shape discrimination method may significantly reduce the fast neutron and the NC atmospheric neutrino events \cite{JUNO:2015zny,JUNO:2022lpc}. In the following, we consider the pulse-shape discrimination method (see Fig.~6 of Ref.~\cite{JUNO:2022lpc} and Table.~5-1 of Ref.~\cite{JUNO:2015zny} for the efficiency of the signals and the backgrounds.) At $E_{e^+}\gtrsim 30\ {\rm MeV}$, the events of atmospheric neutrinos through the CC interactions are the dominant backgrounds.

The expected event rate through the IBD process at JUNO is calculated by
\begin{align}
\frac{dN}{dE_{e^+}}=\epsilon N_t T \int dE_\nu \frac{d\sigma_{\rm IBD}}{dE_{e^+}}\frac{d\Phi_\nu}{dE_\nu},
\end{align}
where $T$ is the exposure time, and the total number of targets (free protons) with the fiducial volume of 17 kton is $N_t=1.2\times 10^{33}$. As for the detector efficiency for the signals, the estimated result is given in Fig.~6 of \cite{JUNO:2022lpc} with the energy range of $12\,{\rm MeV} \leq E_{e^{+}}\leq 30 {\rm MeV}$, and it shows that the efficiency is larger than $0.6$ in the given energy range. Therefore, we set it to be $\epsilon=0.6$ to proceed discussion conservatively.

We use the differential IBD cross section $d\sigma_{\rm IBD}/dE_{e^+}$ from Ref.~\cite{Strumia:2003zx}. As long as $E_\nu\gtrsim (m_n-m_p)$, the range of the integration for neutrino energy is given by \cite{Strumia:2003zx}
\begin{align}
    E_e+\delta \leq E_\nu \leq \frac{E_e+\delta}{1-2(E_{e^+}+\delta)/m_p},
\end{align}
where
\begin{align}
    \delta=\frac{m_n^2-m_p^2-m_e^2}{2m_p},
\end{align}
with the neutron mass $m_n$, the proton mass $m_p$, and the electron mass $m_e$.
Then, the total event rate is expected to be
\begin{align}
    N=\int^{E_{e^+}^{\rm max}}_{E_{e^+}^{\rm min}}dE_{e^+}\frac{dN}{dE_{e^+}},
\end{align}
where $E_{e^+}^{\rm max}$ and $E_{e^+}^{\rm min}$ are given by Eq.~(12) in Ref.~\cite{Strumia:2003zx}.
The energy resolution of the detector is expected to be $\delta_E/E_{e^+}=0.03/\sqrt{(E_{e^+}/{\rm MeV})}$ \cite{JUNO:2015zny}, which is much smaller than the width of the emitted electron energy through the IBD process $\Delta E_{e^+}=E_{e^+}^{\rm max}-E_{e^+}^{\rm min}$ for the energy region of our interest. Thus, we safely ignore the smearing effect of the detector energy resolution.
We have also confirmed that the following results are the same even if the energy resolution is taken into account as a Gaussian profile.

The differential flux of the DSNB for $\bar{\nu}_e$ can be parameterized as
\begin{align}
    \frac{d\Phi^{\rm DSNB}_{\bar{\nu}_e}}{dE_\nu}=\int^{z_{\rm max}}_0 R_{\rm SN}(z)\frac{dN_{\bar{\nu}_e}^{\rm DSNB}}{dE_\nu'}(E_\nu')(1+z)\left|\frac{dt}{dz}\right|dz,
\end{align}
where $z$ is the redshift, $E_\nu'=E_\nu(1+z)$ and $|dt/dz|\simeq H_0(1+z)[\Omega_m(1+z)^3+\Omega_\Lambda]^{1/2}$.
$H_0\simeq 70\ {\rm km\ sec^{-1}\  Mpc^{-1}}$, $\Omega_m\simeq 0.3$ and $\Omega_\Lambda\simeq 0.7$ are the present Hubble parameter, the normalized energy densities of matter and cosmological constant, respectively \cite{Planck:2018vyg}. 
$dN_{\bar{\nu}_e}^{\rm DSNB}/dE_\nu$ is the mean neutrino spectrum from one supernova (SN) for a population of progenitors by weighting each progenitor. 
We assume $z_{\rm max}=5$ to take into account almost all the past SNe. The SN rate $R_{\rm SN}(z)$ is given by
\begin{align}
    R_{\rm SN}(z)=\dot{\rho}^\ast(z)\frac{\int_{8M_{\odot}}^{100M_\odot} dM \psi(M)}{\int_{0.1M_{\odot}}^{100M_\odot} dM M \psi(M)},
    \label{RSN}
\end{align}
where $\dot{\rho}^\ast(z)$ is the star-formulation rate and $\psi(M)=dn/dM$ is the initial mass function (IMF). We assume that the IMF follows the Salpeter law, $\psi(M)\propto M^{-2.35}$. We also assume the following parametrization of the star-formation rate,
\begin{align}
    \dot{\rho}^\ast(z) \propto \left[(1+z)^{p_1k}+\left(\frac{1+z}{5000} \right)^{p_2k}+\left( \frac{1+z}{9} \right)^{p_3k} \right]^{1/k}
\end{align}
with $k=-10,\ p_1=3.4,\ p_2=-0.3$, and $p_3=-3.5$. The SN rate is normalized as $R_{\rm SN}(0)=(1.25\pm 0.5)\times 10^{-4}\ {\rm Mpc}^{-3}\ {\rm yr^{-1}}$ \cite{Lien:2010yb}. 

The mean neutrino spectrum per one SN, $dN_{\bar{\nu}_e}^{\rm DSNB}/dE_\nu$, can be calculated by numerical simulations \cite{Horiuchi:2017qja,Kresse:2020nto}.
We take the mean neutrino spectrum from Fig.~6 of Ref.~\cite{Horiuchi:2017qja}, assuming the progenitor star collapsing into $83\%$ neutron star and $17\%$ black hole.  
In the stellar envelope, flavor conversion effects are enhanced by their interactions with electrons in the medium called the Mikheev-Smirnov-Wolfstein (MSW) effect \cite{Wolfenstein:1977ue,Mikheyev:1985zog,Mikheev:1986if}\footnote{We neglect possible flavor conversion effects induced by neutrino-self interactions in high density region which are currently under study (see, e.g., Ref.~\cite{Airen:2018nvp}).}. Taking into account this effect, the approximate analytical formula of the mean neutrino spectrum at the Earth is given by \cite{Dighe:1999bi,Lu:2016ipr}, assuming the element of the Pontecorvo-Maki-Nakagawa-Sakata matrix, $U_{e3}$, to be the zero, that is, $U_{e3}=0$,
\begin{align}
    \frac{d\Phi_{\nu_e}^{\rm DSNB}}{dE_\nu}&=\frac{d\Phi_{\nu_x}^{\rm DSNB,0}}{dE_\nu},\ \ \ \ 
    \frac{d\Phi_{\bar{\nu}_e}^{\rm DSNB}}{dE_\nu}=c_{12}^2\left(\frac{d\Phi_{\bar{\nu}_e}^{\rm DSNB,0}}{dE_\nu}-\frac{d\Phi_{\nu_x}^{\rm DSNB,0}}{dE_\nu}\right)+\frac{d\Phi_{\nu_x}^{\rm DSNB,0}}{dE_\nu},
\end{align}
where $c_{12}^2=\cos^2\theta_{12}\simeq 0.7$ \cite{Esteban:2020cvm,deSalas:2020pgw} and 
the suffix of ``0'' denotes the initial flux.
For definiteness, we consider only the normal ordering in the neutrino masses since the DSNB flux is almost the same both in the normal and inverted ordering (e.g., see  Ref.~\cite{Moller:2018kpn}).

\begin{figure}
    \centering
    \includegraphics[width = 10cm]{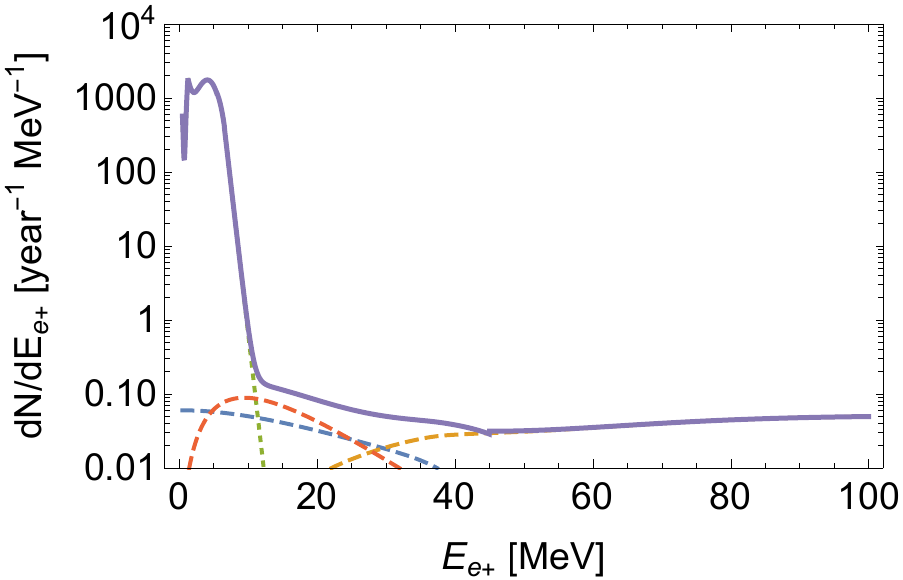}
    \caption{Expected backgrounds in JUNO during 1 year of data-taking. Dashed curves show individual contributions of the three different types of backgrounds: DSNB (red)\cite{Ando:2004sb,Totani:1997vj}, NC atmospheric neutrinos (blue)\cite{JUNO:2015zny}, and CC atmospheric neutrinos (orange)\cite{Battistoni:2005pd}, anticipated to be detected in the fiducial volume of $17$ kton. The dotted green curve represents the reactor neutrino spectrum, anticipated to be detected in $18.35$ kton\cite{JUNO:2015zny}. The solid purple curve represents the total of those backgrounds.}
    \label{fig:bgJUNO}
\end{figure}

In Fig.~\ref{fig:bgJUNO}, we show the event rates for the relevant backgrounds. For the reactor neutrino background, we use the event rate from Fig.~8-4 in Ref.~\cite{JUNO:2015zny}, anticipated to be detected in $18.35$ kton of the detector volume, which is larger than the fiducial volume we are considering. We take them into account without rescaling to the fiducial volume of $17$ kton in order to estimate the bound on the DM annihilation and the decay conservatively. Another possible background might come from ${}^9$Li and ${}^8$He. However, there are still large uncertainties and, in fact, a recent work \cite{Jollet:2019syr} suggests that such background appears in a lower energy region than previously considered. This implies that, even if we would take them into account, the results will not change much in our case. Hence, we do not consider them in the present work.
For the background of the NC atmospheric neutrino events, we take the event rate from the right panel of Fig.~5-2 of Ref.~\cite{JUNO:2015zny}.
For the background of CC atmospheric $\bar{\nu}_e$ events, we also take the atmospheric $\bar{\nu}_e$ flux from the result of the FLUKA simulations \cite{Battistoni:2005pd}. Since the location of JUNO is approximately the same latitude as that of SK, we make use of the same value in Table~3 for SK of Ref.~\cite{Battistoni:2005pd} in the FLUKA simulations.

\subsection{Discovery potential for neutrino lines from MeV DM annihilation and decay}
\label{sec3.2}

To obtain the limits on the annihilation cross section and the lifetime of DM, we perform the analogous analysis proposed in Ref.~\cite{Moller:2018kpn}.
We define the anticipated observed numbers of each background (BG) source and DM signal within an energy bin $\Delta$ centered at $E_{i}$,
\begin{align}
    N_{i}^{\rm BG} 
    &= T\int_{E_{i}-\Delta/2}^{E_{i}+\Delta/2}
    dE_{e^+}\left(\frac{dN^{\mathrm{DSNB}}}{dE_{e^+}}+\frac{dN^{\mathrm{atm,NC}}}{dE_{e^+}}+\frac{dN^{\mathrm{atm,CC}}}{dE_{e^+}}+\frac{dN^{\mathrm{reactor}}}{dE_{e^+}}\right), \label{NBG} \\
    N_{i}^{\rm DM} 
    &= T\int_{E_{i}-\Delta/2}^{E_{i}+\Delta/2}
    dE_{e^+}\frac{dN^{\mathrm{DM}}}{dE_{e^+}},
    \label{NDM}
\end{align}
where $T$ is the exposure time.
The total numbers of expected observable signals per an energy bin are given by
\begin{equation}
    N_{i}(\bm{\theta}) = N_{i}^{\rm BG} (\bm{\theta})+N_{i}^{\rm DM} (\bm{\theta}),\quad
    N_{i}^{\rm obs}(\hat{\bm{\theta}}) = N_{i}^{\rm BG}(\hat{\bm{\theta}}),
\end{equation}
where $\bm{\theta} = (\langle \sigma v \rangle)$ or $(\tau)$ in our discussion of the annihilation or the decay of DM respectively, and parameters with hats are the fiducial values. We have confirmed that, even if we take into account the uncertainties of DSNB, which is mainly that of the SN rate $R_{\rm SN}(z)$ of $\sim 30\%$, and atmospheric neutrino spectra of $\sim 30\%$ \cite{Battistoni:2005pd}, the constraints on the DM lifetime do not change significantly. Then, in the present work, we do not marginalize the $\chi^{2}$-function with respect to such background uncertainties. The $\chi^{2}$-function is defined as 
\begin{equation}
    \chi^{2}(\bm{\theta})
    =-2 \sum_{i}\ln\frac{L_{0,E_{i}}}{L_{1,E_{i}}},
\end{equation}
where the sum is taken over the all of energy bins. The likelihood functions are estimated as
\begin{align}
    L_{0,E_{i}}&\simeq
    P_{i}[\lambda = N_{i}^{\rm obs}(\hat{\bm{\theta}}),\,k=N_{i}(\bm{\theta})],\\
    L_{1,E_{i}}&\simeq
    P_{i}[\lambda = N_{i}(\bm{\theta}),\,k=N_{i}(\bm{\theta})],
\end{align}
where $P_{i} = \lambda^{k}\exp(-\lambda)/k!$ is a Poisson distribution with a mean value $\lambda$ and a observed number $k$ at the $i$-th energy bin.
The parameters $\bm{\theta}$ are ruled out at $n\sigma$ when $\sqrt{\chi^{2}(\bm{\theta})}>n$. 
In the following, we take the energy bin as $\Delta= 0.1\, {\rm MeV}$ in Eqs.~(\ref{NBG}) and (\ref{NDM}), which is comparable to the difference between the maximum and minimum positron energies for IBD in the range $m_{\chi}=\mathcal{O}(10\,\mathrm{MeV})$.

\begin{figure}
    \centering
    \includegraphics[width = 15cm]{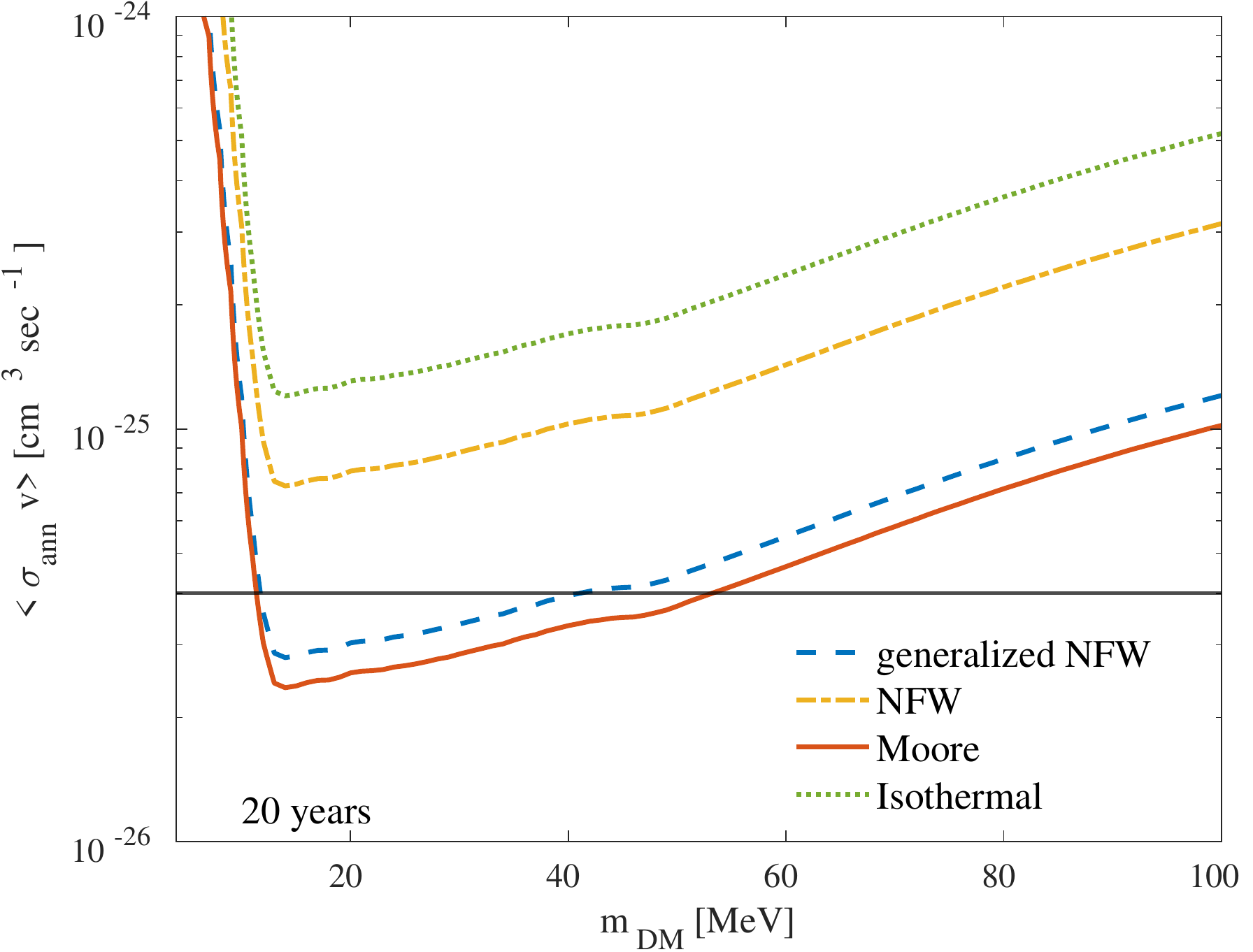}
    \caption{$90\%$ C.L.\ anticipated bound on the DM annihilation cross section from the whole Milky Way with IBD detection channel in JUNO. The dashed blue curve, dot-dashed yellow curve, solid red curve and dotted green curve correspond to the constraints obtained from $20$ years of data-taking, assuming DM profile as generalized NFW, NFW, Moore and Isothermal, respectively. The solid black line corresponds to the annihilation cross section of $\langle \sigma v \rangle = 4 \times 10^{-26}\ {\rm cm^3\  sec^{-1}}$ for the thermal production mechanism of DM.}
    \label{fig:constraint_annihilation}
\end{figure}

\begin{figure}
    \centering
    \includegraphics[width = 15cm]{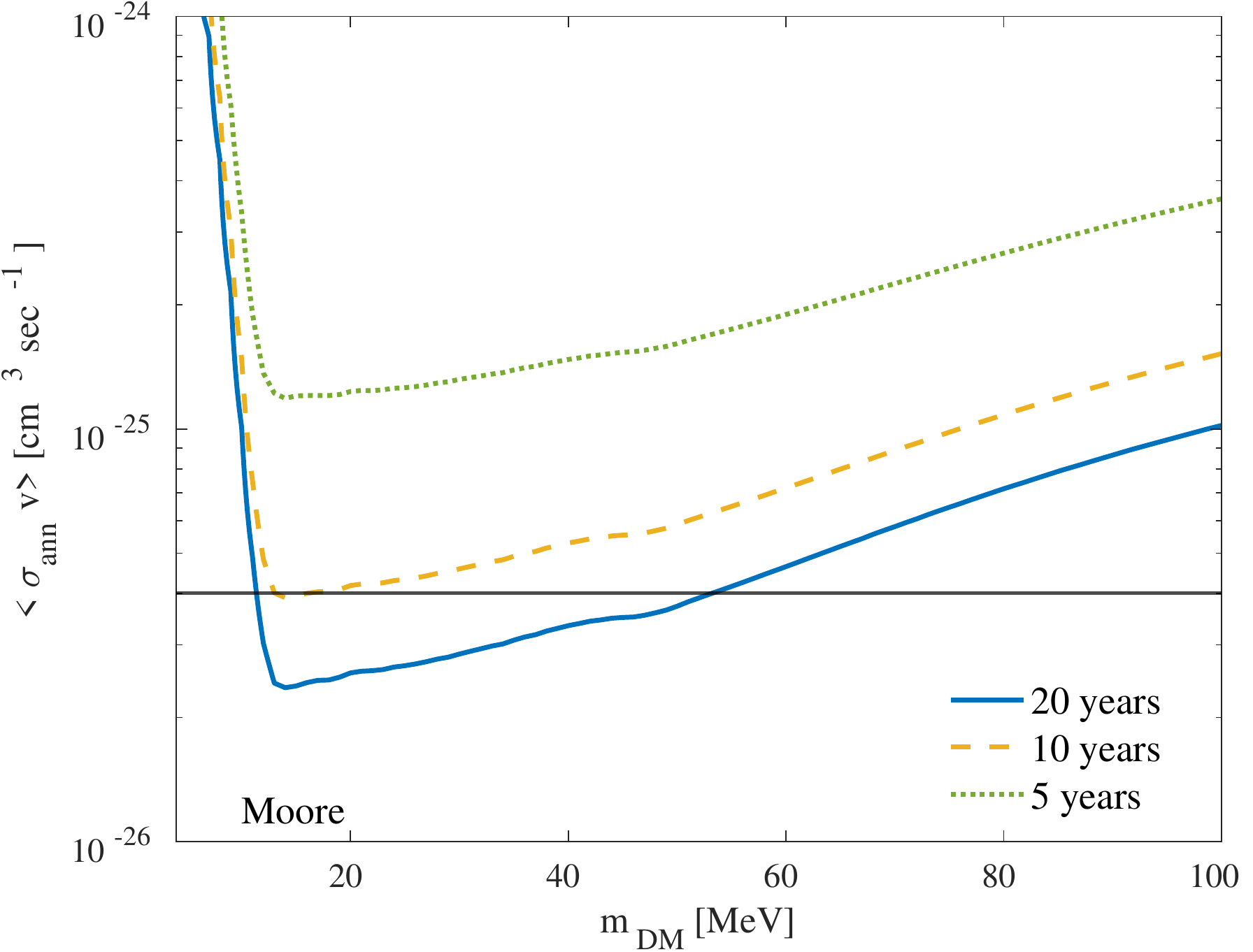}
    \caption{$90\%$ C.L.\ anticipated bound on the DM annihilation cross section from the whole Milky Way with IBD detection channel in JUNO. The solid blue curve, dashed yellow curve, and dotted green curve correspond to the constraints obtained from $20,\,10,\,5$ years of data-taking, respectively, assuming the Moore profile. The solid black line corresponds to the annihilation cross section of $\langle \sigma v \rangle = 4 \times 10^{-26}\ {\rm cm^3\  sec^{-1}}$ for the thermal production mechanism of DM.}
    \label{fig:constraint_annihilation_moore}
\end{figure}

\begin{figure}
    \centering
    \includegraphics[width = 15cm]{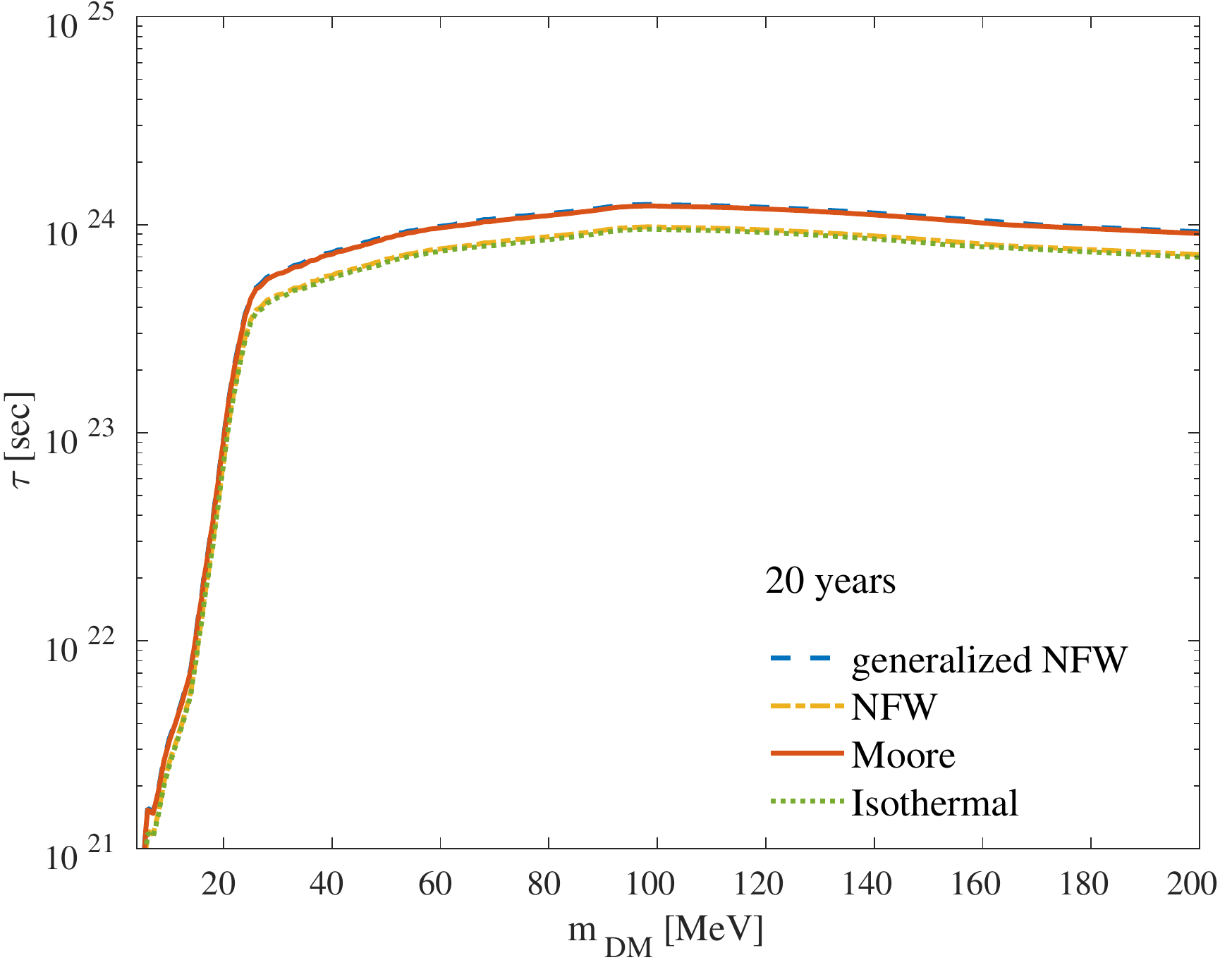}
    \caption{$90\%$ C.L.\ anticipated bound on the DM lifetime from the whole Milky Way with IBD detection channel in JUNO. The dashed blue curve, dot-dashed yellow curve, solid red curve and dotted green curve correspond to the constraints obtained from $20$ years of data-taking, assuming DM profile as generalized NFW, NFW, Moore and Isothermal, respectively.}
    \label{fig:constraint_decay}
\end{figure}

\begin{figure}
    \centering
    \includegraphics[width = 15cm]{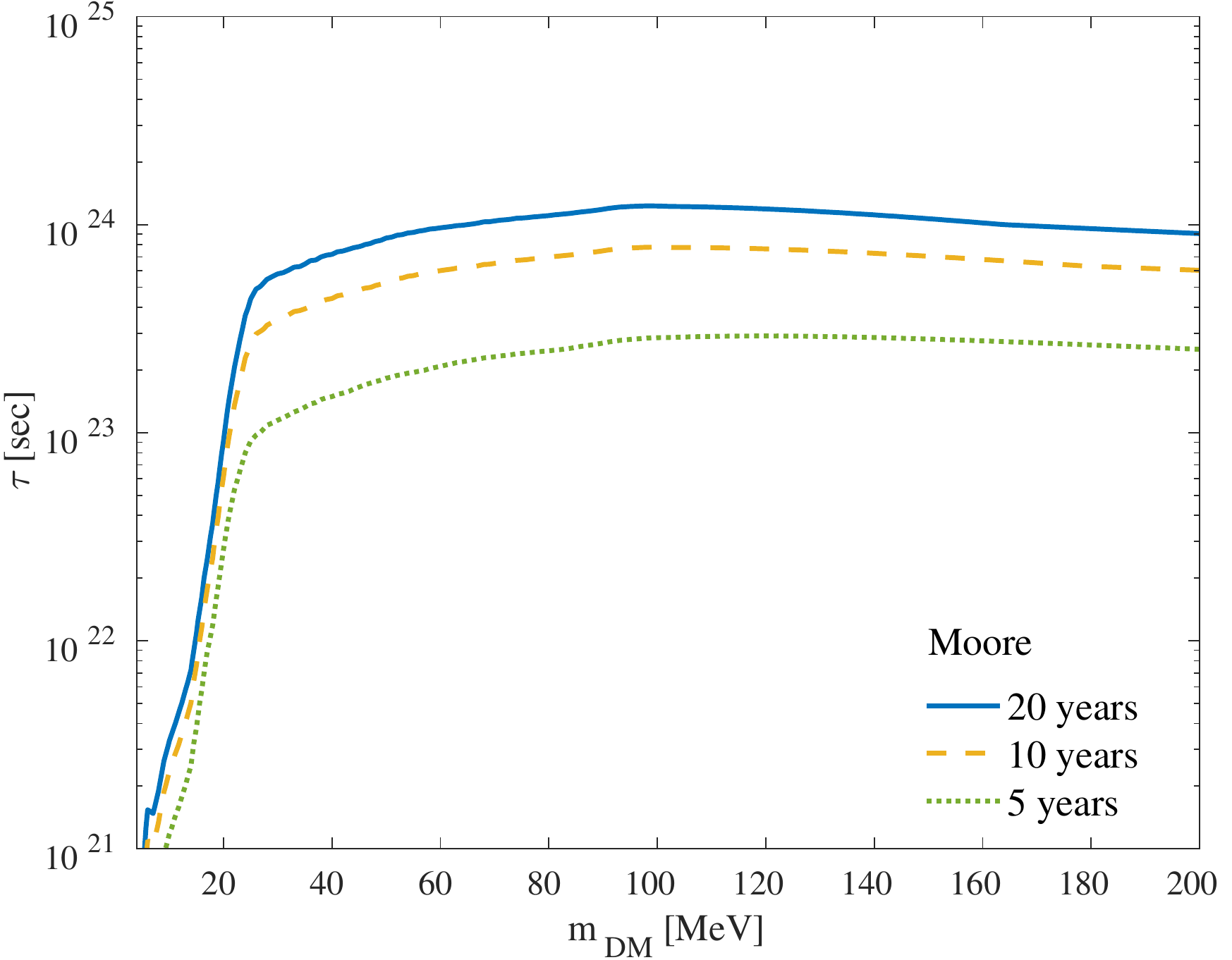}
    \caption{$90\%$ C.L.\ anticipated bound on the DM lifetime from the whole Milky Way with IBD detection channel in JUNO. The solid blue curve, dashed yellow curve, and dotted green curve correspond to the constraints obtained from $20,\,10,\,5$ years of data-taking, respectively, assuming the Moore profile.}
    \label{fig:constraint_decay_moore}
\end{figure}

Figs.\ \ref{fig:constraint_annihilation} and \ref{fig:constraint_decay} show the 90\% C.L.\ constraints on the annihilation cross section and the lifetime of DM from JUNO experiment respectively, with the energy bin $\Delta = 0.1\,{\rm MeV}$ and 20 years data-taking. Four curves correspond to the generalized NFW (dashed blue curve), NFW (dot-dashed yellow curve), Moore (solid red curve), and Isothermal (dotted green curve) profile, respectively. Figs.\ \ref{fig:constraint_annihilation_moore} and \ref{fig:constraint_decay_moore} also show the 90\% C.L.\ constraints on the annihilation cross section and the lifetime of DM from JUNO experiment, respectively, with the total time exposure of $5$ years (dotted green curve), $10$ years (dashed yellow curve) and $20$ years (solid blue curve), assuming the Moore profile.

As for the annihilation cross section, the most stringent bound is obtained at $m_{\chi} \simeq 15\,{\rm MeV}$, where the background from reactor neutrinos is significantly suppressed. The constraint above $m_{\chi} \simeq 15\,{\rm MeV}$ gets weaker as the mass of DM particle gets larger, which shows a slightly stronger dependence of the DM mass than the constraint given in Fig. 2 of Ref.\ \cite{Arguelles:2019ouk}.
This might come from the facts that the neutrino flux from the DM annihilation is suppressed by $m_{\chi}^{-2}$, and that the emitted positron signal with higher energy through the IBD process is more smeared, resulting in the smaller ratio between the signal and the approximately constant background. The comparison with Ref.\ \cite{Klop:2018ltd} will be given in Appendix~\ref{appa} because Ref.\ \cite{Klop:2018ltd} includes the effects of $\mathrm{^{12}C}$. We find that the constraint on the annihilation cross section into neutrinos might reach the value of the thermal annihilation cross section of DM of $\langle \sigma v \rangle \sim 4 \times 10^{-26}\, {\rm cm^3\, sec^{-1}}$ after $20$ years of data-taking from $m_{\chi}\simeq 15\, {\rm MeV}$ to $m_{\chi}\simeq 40$ and 50$\, {\rm MeV}$, assuming the generalized NFW and the Moore profile, respectively.

As for the lifetime, the suppression of the neutrino number flux from the DM decay with the larger mass of DM particles is $m_{\chi}^{-1}$ as given in Eq.\ \eqref{eq:DM_decay_flux}, which is smaller than the annihilation case. Then, the most stringent bound is obtained at 
$m_{\chi} \simeq 100\,{\rm MeV}$, where the background reaches its minimum, and the mass dependence on the bound in the larger mass region gets softer compared to the annihilation case.
We find that the lower bound after $20$ years of data-taking is $\tau \gtrsim 1 \times 10^{24}\,{\rm sec}$, almost independently of the mass of DM particle in the mass region $40\,\mathrm{MeV}<m_{\chi}<200\,\mathrm{MeV}$,
while the previous constraint obtained from Super-Kamiokande \cite{Palomares-Ruiz:2007egs} has two explicit peaks at $m_{\chi}\simeq 50\,\mathrm{MeV}$ and $m_{\chi}\simeq140\,\mathrm{MeV}$, whose corresponding lower bounds are $\tau \gtrsim \num{4e23}\,{\rm sec}$ and $\tau \gtrsim\num{7e23}\,{\rm sec}$, respectively.


\section{Conclusions}
\label{sec7}

In the present paper, we have considered annihilation and decay of DM particles in the whole Milky Way into a pair of neutrino-antineutrino, and we have shown the future constraints on the annihilation cross section and the lifetime of MeV-scale DM particles in JUNO experiment. A similar analysis has already been done in Refs.~\cite{Klop:2018ltd,Arguelles:2019ouk} for annihilation. We extend the analysis by taking into account longer data-taking, wider mass range, and different DM profiles. Furthermore, we have included analysis for the DM decay and revealed the sensitivity in JUNO.
The main forecasts we have found are summarised as follows:
\begin{itemize}
    \item For the annihilation of DM, JUNO will succeed in constraining the cross section as strong as $\num{4e-26}\,{\rm cm^{3}\, sec^{-1}}$ for the mass of a DM particle from $m_{\chi} \simeq 15\,{\rm MeV}$ to $m_{\chi} \simeq 40\ (50)\,{\rm MeV}$, at $90\,\%$ C.L.\ with $20$ years of data-taking in the fiducial volume $17$ kton, assuming the generalized NFW (Moore) profile, respectively. 
    We have also obtained the bound in the mass region above $50\,{\rm MeV}$ considering CC atmospheric $\bar{\nu}_{e}$ background.
    The constraint on the cross section gets weaker as the mass of DM gets larger, especially in the region where the dominant background comes from CC atmospheric $\bar{\nu}_{e}$, whose capture rate barely depends on the positron energy. This might come from the facts that the neutrino flux from the DM annihilation process is suppressed in proportion to $m_{\chi}^{-2}$, and that the emitted positron signal with larger energy is more smeared through the IBD process.
    
    \item For the decay of DM, JUNO will succeed in updating the current constraint from SK \cite{Palomares-Ruiz:2007egs} and in constraining the lifetime as strong as $\num{1e24}\,{\rm sec}$ for the mass of a DM particle $m_{\chi} \simeq 100\,{\rm MeV}$, at $90\,\%$ C.L.\ with $20$ years of data-taking in the fiducial volume $17$ kton.
    Since the neutrino flux from the DM decay process is proportional to $m_{\chi}^{-1}$, which is less suppressed than that from the DM annihilation process, the constraint on the lifetime with larger masses of DM does not necessarily get weaker. Therefore, the strongest bound is not obtained at the mass $m_{\chi} \simeq 30\,{\rm MeV}$, where the reactor neutrino background is significantly suppressed, but at the mass $m_{\chi} \simeq 100\,{\rm MeV}$, where the total background gets its minimum. 
\end{itemize}

The current lower limit on the thermal annihilation cross section into neutrinos is $m_{\chi} \lesssim 3.7\ {\rm MeV}$ from BBN and CMB observations \cite{Boehm:2013jpa,Nollett:2013pwa,Nollett:2014lwa,Escudero:2018mvt,Sabti:2019mhn}. Upcoming Hyper-Kamiokande and CMB-S4 experiments are also expected to improve the sensitivity of annihilation and decay into neutrinos at the MeV to sub-GeV range of DM mass \cite{Olivares-DelCampo:2018pdl,Bell:2020rkw, Escudero:2018mvt,Sabti:2019mhn}.
The combination of JUNO and these experiments will be complementary and comprehensive to probe DM annihilation and decay into neutrinos.
For DM annihilation into neutrinos, CMB-S4 observation is sensitive to the annihilation cross section for the DM mass up to 10-15$\, {\rm MeV}$ \cite{Escudero:2018mvt,Sabti:2019mhn} and Hyper-Kamiokande is sensitive to that around the DM mass of $20\, {\rm MeV}$. In our analysis, JUNO is also sensitive to that around the DM mass of $15\, {\rm MeV}$, which can indeed be complementary. 

In summary, we have discussed the model-independent constraints on annihilation and decay of MeV-scale DM into a pair of neutrino-antineutrino, and we have shown that JUNO will improve the current constraints in both processes, and could test the thermal production mechanism of DM particles after $20$ years of data-taking for some range of the DM mass.


\section*{Acknowledgments}
We would like to thank Shin'ichiro Ando for the correspondence on Ref.\ \cite{Klop:2018ltd}. KA is supported by IBS under the project code, IBS-R018-D1. GL is supported by INFN and by MIUR. MY acknowledges financial support from JSPS Grant-in-Aid for Scientific Research No. JP18K18764, JP21H01080, JP21H00069.


\appendix

\section{Discovery potential including the charged current neutrino interaction with $^{12}\mathrm{C}$}
\label{appa}

In this appendix, we consider the effect of the subdominant CC neutrino interactions with $^{12}\mathrm{C}$ on the discovery potential of JUNO for the neutrino lines from DM. 
In Section~\ref{sec3}, we considered only the IBD channel to detect the DM signal since we might distinguish the IBD channel from the other channels by neutron tagging. In case that we include the channels of the CC interactions for $\nu_e$ and $\bar{\nu}_e$ with $^{12}\mathrm{C}$, both the neutrino signal from DM and the background from the atmospheric neutrinos are increased, which might have a chance to enhance or to suppress the sensitivity of JUNO. However, we find that the sensitivity of JUNO becomes slightly weaker due to the increase of the background in this case.

To calculate the differential cross sections for the CC interactions of $\nu_e$ and $\bar{\nu}_e$ with $^{12}\mathrm{C}$ (bounded nucleons), we use a relativistic Fermi gas model \cite{Smith:1972xh} with a Fermi momentum of $221\, {\rm MeV}$ and a binding energy of $25\, {\rm MeV}$, following a similar analysis of oxygen in SK in Ref.~\cite{Palomares-Ruiz:2007trf}. The event rates for the signal and the background by taking the CC interactions with $^{12}\mathrm{C}$ into account are calculated in the same manner as was done in Section~\ref{sec3.1}. In these event rates, we also neglect the small smearing effect of the energy resolution of JUNO because we have confirmed that the results remain almost unchanged even after taking into account the energy resolution with a Gaussian profile.

\begin{figure}
    \centering
    \includegraphics[width = 10cm]{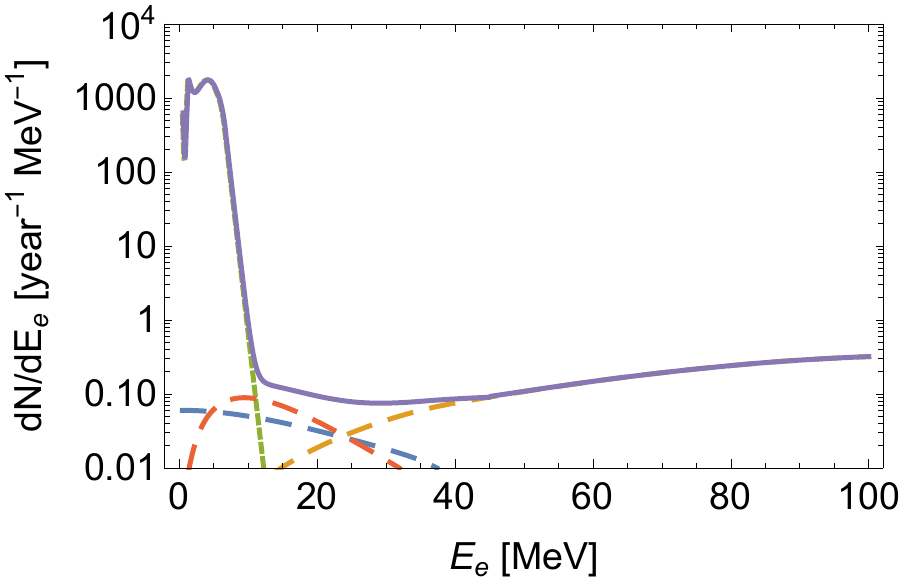}
    \caption{Expected backgrounds in JUNO during 1 year of data-taking, including the CC neutrino interaction with $^{12}\mathrm{C}$. Dashed curves show individual contributions of the three different types of backgrounds: DSNB (red)\cite{Ando:2004sb,Totani:1997vj}, NC atmospheric neutrinos (blue)\cite{JUNO:2015zny}, and CC atmospheric neutrinos (orange)\cite{Battistoni:2005pd}, anticipated to be detected in the fiducial volume of $17$ kton. The dotted green curve represents the reactor neutrino spectrum, anticipated to be detected in $18.35$ kton\cite{JUNO:2015zny}. The solid purple curve represents the total of those backgrounds.}
    \label{fig:bgJUNO_carbon}
\end{figure}

\begin{figure}
    \centering
    \includegraphics[width = 15cm]{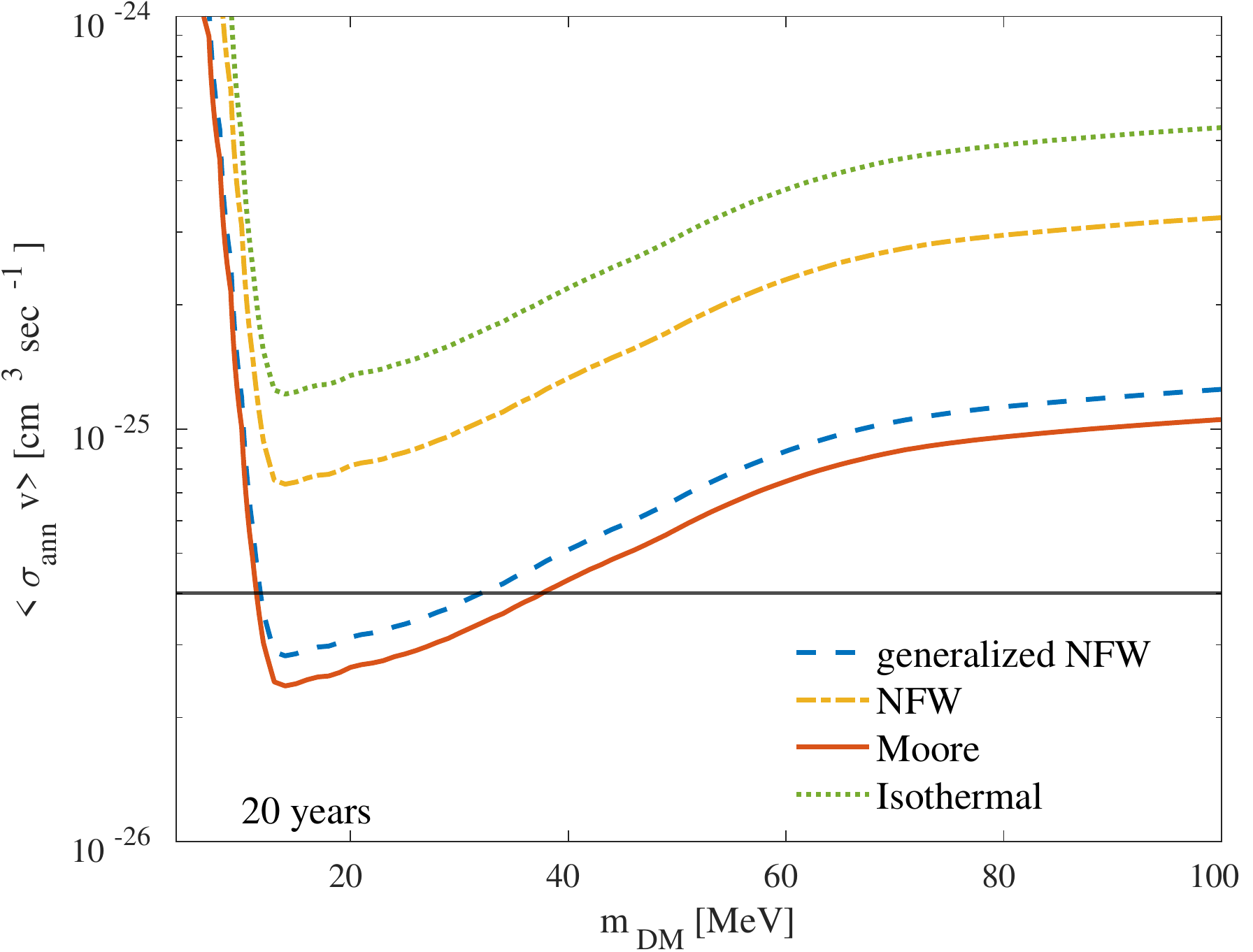}
    \caption{$90\%$ C.L.\ anticipated bound on the DM annihilation cross section from the whole Milky Way with IBD and CC interaction with $^{12}$C detection channel in JUNO. The dashed blue curve, dot-dashed yellow curve, solid red curve and dotted green curve correspond to the constraints obtained from $20$ years of data-taking, assuming DM profile as generalized NFW, NFW, Moore and Isothermal, respectively. The solid black line corresponds to the annihilation cross section of $\langle \sigma v \rangle = 4 \times 10^{-26}\ {\rm cm^3\  sec^{-1}}$ for the thermal production mechanism of DM.}
    \label{fig:constraint_annihilation_c}
\end{figure}

\begin{figure}
    \centering
    \includegraphics[width = 15cm]{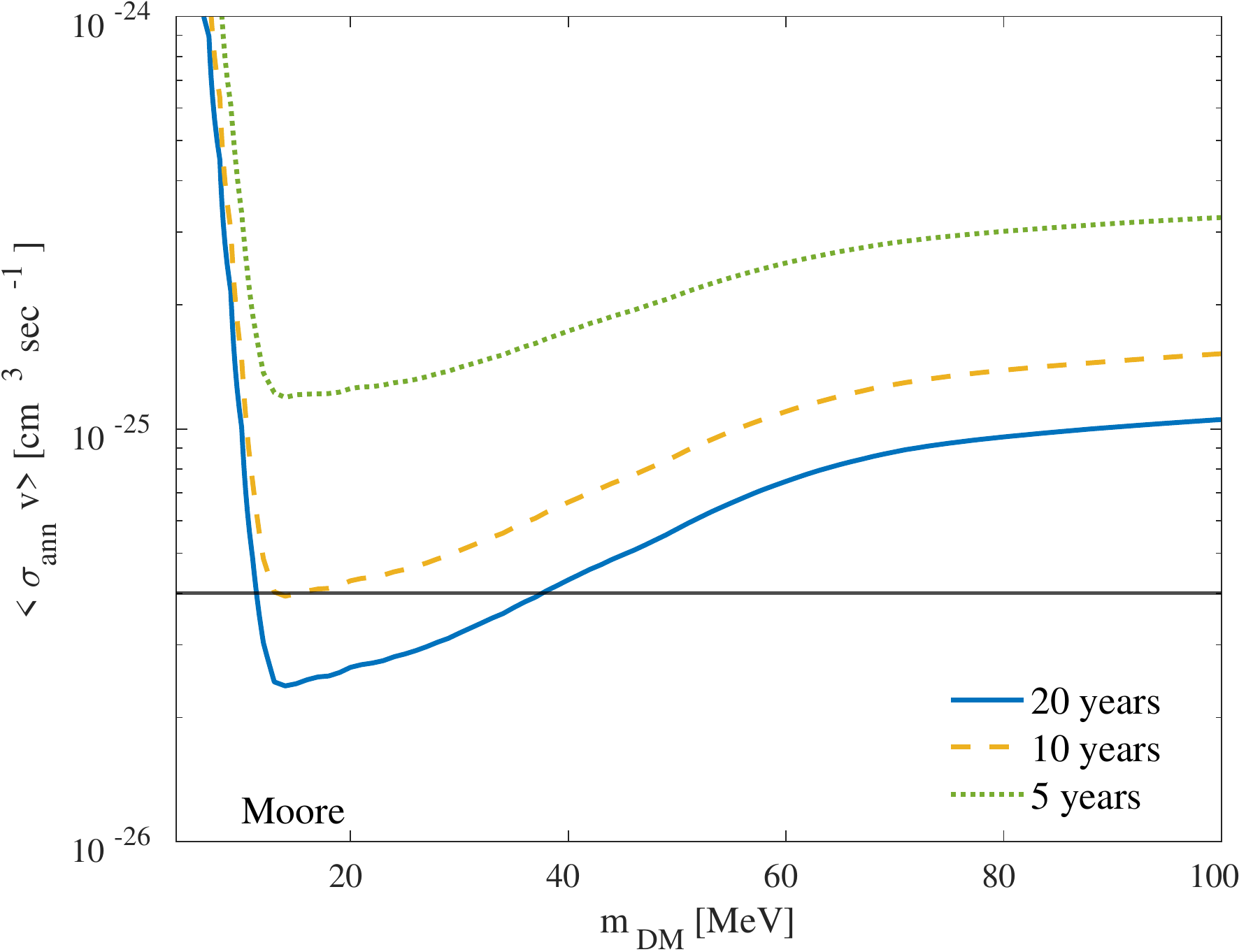}
    \caption{$90\%$ C.L.\ anticipated bound on the DM annihilation cross section from the whole Milky Way with IBD and CC interaction with $^{12}$C detection channel in JUNO. The solid blue curve, dashed yellow curve, and dotted green curve correspond to the constraints obtained from $20,\,10,\,5$ years of data-taking, respectively, assuming the Moore profile. The solid black line corresponds to the annihilation cross section of $\langle \sigma v \rangle = 4 \times 10^{-26}\ {\rm cm^3\  sec^{-1}}$ for the thermal production mechanism of DM.}
    \label{fig:constraint_annihilation_moore_c}
\end{figure}

\begin{figure}
    \centering
    \includegraphics[width = 15cm]{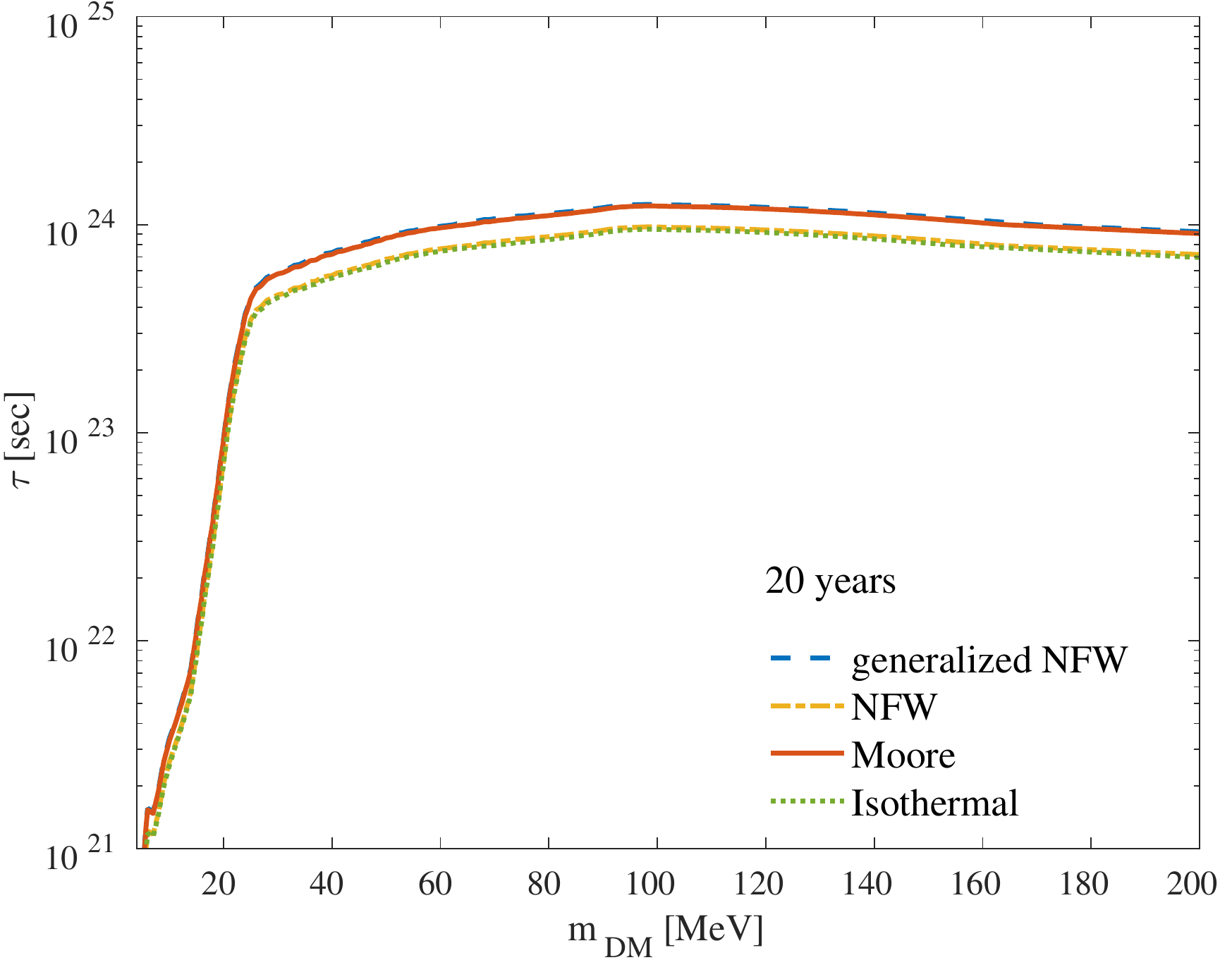}
    \caption{$90\%$ C.L.\ anticipated bound on the DM lifetime from the whole Milky Way with IBD and CC interaction with $^{12}$C detection channel in JUNO. The dashed blue curve, dot-dashed yellow curve, solid red curve and dotted green curve correspond to the constraints obtained from $20$ years of data-taking, assuming DM profile as generalized NFW, NFW, Moore and Isothermal, respectively.}
    \label{fig:constraint_decay_c}
\end{figure}

\begin{figure}
    \centering
    \includegraphics[width = 15cm]{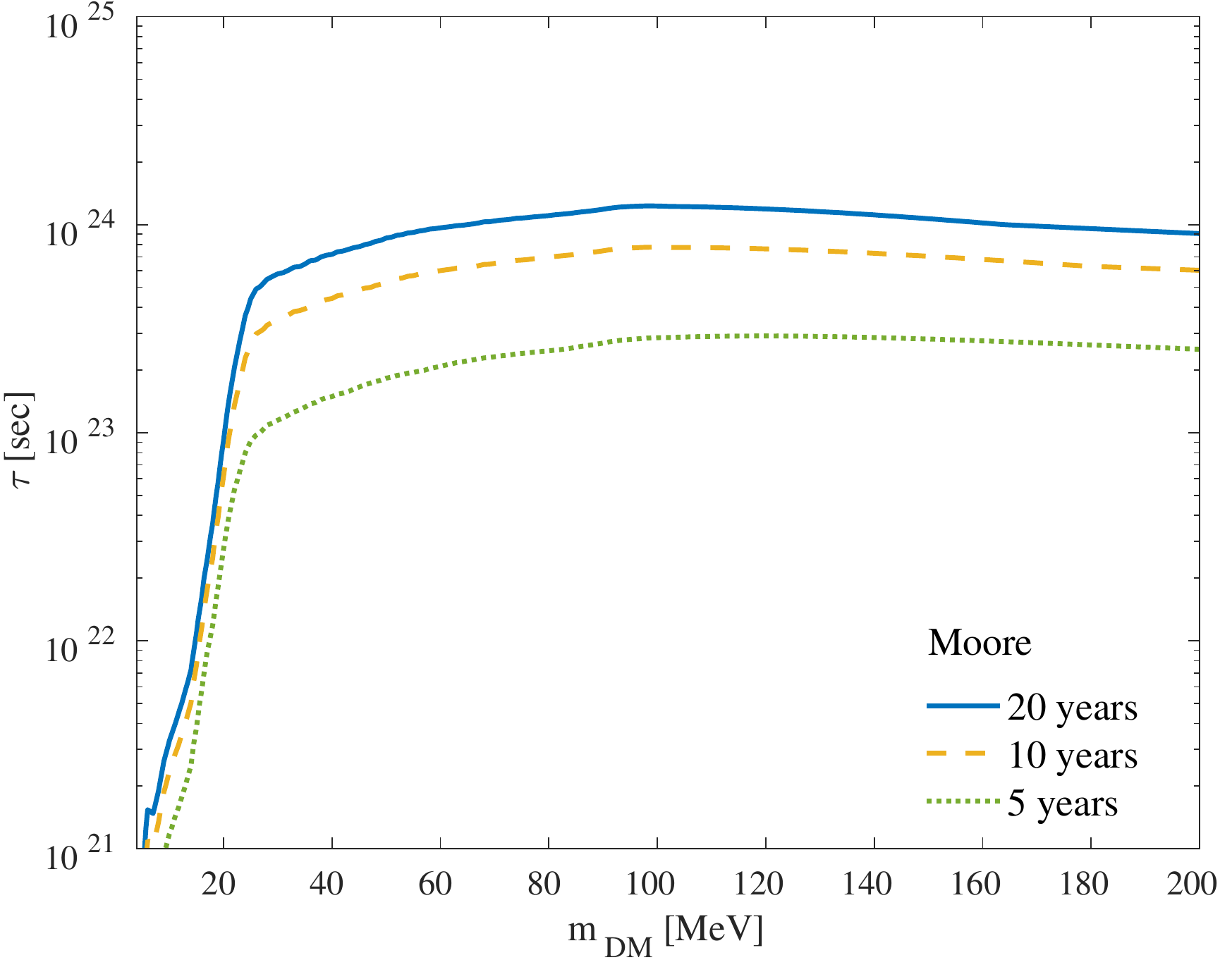}
    \caption{$90\%$ C.L.\ anticipated bound on the DM lifetime from the whole Milky Way with IBD and CC interaction with $^{12}$C detection channel in JUNO. The solid blue curve, dashed yellow curve, and dotted green curve correspond to the constraints obtained from $20,\,10,\,5$ years of data-taking, respectively, assuming the Moore profile.}
    \label{fig:constraint_decay_moore_c}
\end{figure}

In Fig.~\ref{fig:bgJUNO_carbon}, we show the event rates for the relevant backgrounds, including the background from the CC interactions for atmospheric $\nu_e$ and $\bar{\nu}_e$ with $^{12}\mathrm{C}$.
Due to the additional CC interaction for atmospheric neutrinos with $^{12}\mathrm{C}$, the dashed orange line (and the solid purple line) in Fig.~\ref{fig:bgJUNO_carbon} is higher than that in Fig.~\ref{fig:bgJUNO}. 
Figs.\ \ref{fig:constraint_annihilation_c} and \ref{fig:constraint_decay_c} show the 90\% C.L.\ constraints on the annihilation cross section and the lifetime of DM for the four DM profiles from IBD and CC interaction with $^{12}$C detection channel in JUNO experiment, respectively, assuming 20 years data-taking.
Figs.~\ref{fig:constraint_annihilation_moore_c} and \ref{fig:constraint_decay_moore_c} also show the 90\% C.L.\ constraints on the annihilation cross section and the lifetime of DM for the four DM profiles from IBD and CC interaction with $^{12}$C detection channel with 5, 10 and 20 years of data-taking, respectively, assuming the Moore profile.

As discussed in Sec.~\ref{sec3.2}, the constraints get weaker as the DM mass gets larger due to the mass dependence on the DM flux (see Eqs.~\eqref{eq:DM_anni_flux} and \eqref{eq:DM_decay_flux}). 
Compared with the results in Sec.~\ref{sec3.2}, the constraints are weaker in the DM mass of $\sim30$-$100$ ($60$-$200$)$\, {\rm MeV}$ for the annihilation (decay).
This fact might come from that the background in the relevant region is enhanced by the CC interaction of $^{12}$C for relatively high energy atmospheric neutrinos with $E_{e^\pm} \simeq E_{\nu} -25\,{\rm MeV} $ in the relativistic Fermi gas model, while the DM signals with the above mass are mainly detected by the IBD process with $E_{e^+} \simeq E_{\nu} -1.3\,{\rm MeV}$, since the additional detection channel is not useful to detect line signals due to its large smearing effect. 
For DM mass of $\gtrsim 100\ (200)\, {\rm MeV}$ for the annihilation (decay), the DM signals might be also mainly detected by the CC interaction with $^{12}$C, improving the sensitivity of JUNO.

The obtained constraints on the annihilation cross section are slightly different from those given in Ref.~\cite{Klop:2018ltd}. The difference originates mainly from the different treatments of the cross sections for the IBD process and the CC neutrino interaction with $^{12}$C. In Ref.~\cite{Klop:2018ltd}, they computed the number of capture events of DM signals using the {\it total} cross section by neglecting the difference between the maximum and minimum electron (positron) energies emitted at neutrino (anti-neutrino) detection. In the present paper, we have used the {\it differential} cross section. Under the current situation that the uncertainties coming from the DM profile is larger than the difference of the results from these two different approaches, the treatment taken in Ref.~\cite{Klop:2018ltd} is simpler and justified. However, once the uncertainty of the DM profile would get much smaller in the future, our treatment with the differential cross section to deal with the smearing effect by the width of emitted electrons (positrons) will be more adequate.


\bibliography{reference} 
\bibliographystyle{JHEP}

\end{document}